\documentstyle[aps,prb,multicol]{revtex}

\title{Interaction Effects in a One-Dimensional Constriction}
\author{K.~J. Thomas, J.~T. Nicholls, N.~J. Appleyard, M. Pepper,\\ 
M.~Y. Simmons, D.~R. Mace, W.~R. Tribe, and D.~A. Ritchie}
\address{Cavendish Laboratory, Madingley Road, Cambridge CB3 0HE, UK}
\date{\today}
\begin{document} 
\maketitle 
\begin{abstract} 
We have investigated the transport properties of
one-dimensional (1D) constrictions defined by split-gates in high quality
GaAs/AlGaAs heterostructures. In addition to the usual quantized conductance
plateaus, the equilibrium conductance shows a structure close to
$0.7(2e^2/h)$, and in consolidating our previous work 
[K.~J. Thomas {\it et~al.}, Phys.\ Rev.\ Lett. {\bf 77}, 135 (1996)] 
this {\it 0.7 structure}
has been investigated in a wide range of samples as a function of
temperature, carrier density, in-plane magnetic field $B_{\parallel}$ and
source-drain voltage $V_{sd}$. 
We show that the 0.7 structure is not due to
transmission or resonance effects, nor does it arise from the asymmetry of
the heterojunction in the growth direction. All the 1D subbands show Zeeman
splitting at high $B_{\parallel}$, and in the wide channel limit the
$g$-factor is $\mid g \mid \approx 0.4$, close to that of bulk GaAs. As the
channel is progressively narrowed we measure an exchange-enhanced $g$-factor.
The measurements establish that the 0.7 structure is related to spin, 
and that electron-electron interactions become important 
for the last few conducting 1D subbands.

\end{abstract}
\begin{multicols}{2}
\section{Introduction}\label{s:intro}

When a negative gate voltage is applied to a lithographically defined
split-gate, the underlying two-dimensional electron gas (2DEG) is
electrostatically squeezed into a one-dimensional (1D)
channel.\cite{thornton86} 
In a clean 1D constriction, where the
mean free path is much longer than the effective channel length, the
conductance is quantized\cite{Wharam88,VanWees88} in units of $2e^2/h$; a
result that can be understood as the adiabatic transmission of 1D subbands. 
In an earlier paper\cite{thomas96a} we showed that, in addition to
the usual quantized conductance plateaus, there is also a structure at
$0.7(2e^2/h)$. This so-called {\it 0.7 structure} shows characteristics that
demonstrate the importance of many-body interactions in the limit of a few
conducting 1D subbands. 

As a consequence of electron-electron interactions, a 1D electron gas is
expected to exhibit Tomonaga-Luttinger\cite{tomanaga50,luttinger63} (TL)
liquid behavior rather than Fermi liquid behavior. 
In addition to a TL liquid
there are other possible states of an interacting 1D system,
for example, a 1D Wigner crystal is predicted\cite{glazman92} 
when the 1D electron density is less than the (Bohr\ radius)$^{-1}$. 
It has also been shown\cite{bloch29} 
that at sufficiently low electron densities 
the exchange interactions will 
dominate over the kinetic energy, and a three-dimensional 
electron gas will undergo a transition to a ferromagnetic state.
The increasing importance of the exchange interactions in lower dimensions
is borne out by recent calculations\cite{gold96,wang96}  
that show a similar spontaneous spin 
polarization in a quasi-one dimensional electron gas.

In the light of these ideas we present experimental evidence showing that
electron-electron interactions are important in a ballistic 1D constriction. 
We do not observe TL liquid behavior, but instead we believe there is
evidence for spontaneous spin polarization. We expand upon our earlier
work,\cite{thomas96a} showing results for six different samples. The rest of
this paper is organized as follows. Section~\ref{s:review} gives a
brief review of split-gate devices, before a description of the samples and
measurements in Sec.~\ref{s:exp}. The zero-field gate characteristics as a
function of temperature and 2D carrier density are presented in
Sec.~\ref{s:quantiz}, and measurements in a strong in-plane magnetic field
and with an applied source-drain voltage 
in Secs.~\ref{s:mag} and~\ref{s:sd}, respectively. 
We discuss our results, 
and their relevance to the TL model
in Sec.~\ref{s:disc}.

\section{Review of Split-Gate Devices}
\label{s:review}

Split-gates\cite{thornton86} are a well established\cite{BeenRev} technique
for creating a smooth one-dimensional constriction in a 2DEG. When a negative
voltage $V_g$ is applied to a lithographically defined pair of Schottky 
split-gates above a GaAs/AlGaAs heterostructure, shown in Fig.~\ref{f:1}(a),
the 2DEG is depleted from beneath the gates and a 1D channel is left defined
between them. If the elastic mean free path $l_e$ is much greater than the
width $W$ and length $L$ of the channel, transport through the 1D
constriction is ballistic and the differential conductance,  
$G(V_g) = N (2e^2/h)$, is quantized,\cite{Wharam88,VanWees88}  
where $N$ is the number of transmitted 1D subbands.  
At small negative gate voltages, when a
wide 1D channel is first defined, the lateral confinement potential is best
described by a square well with a width similar to the lithographic
separation $W$ of the split-gates, and an electron density equal to that of
the bulk 2DEG ($n_{2D}$). The carrier density and width of the channel are
progressively reduced as $V_g$ becomes more negative, and when there are only
two or three occupied 1D subbands the electrostatic landscape around the
narrowest part of the constriction can be modeled by a saddle-point
potential.\cite{lmm92} 
Split-gate structures have been used to study electron focussing,\cite{VanHouten88}
non-linear transport\cite{patel91b,kouwen89} and magnetic
depopulation,\cite{berggren86} all of which can be interpreted in a
non-interacting electron picture. 
Recent conductance measurements\cite{thomas96a} of ultra-clean split-gate devices 
exhibit a structure at $0.7(2e^2/h)$ that cannot be explained 
within a non-interacting picture.

In this paper, as well as in recent work,\cite{thomas96a,thomas95} the 1D
constrictions are defined in deep heterostructures where the 2DEG is up to
3000~\AA\ below the sample surface. Using these high purity 2DEGs (with a low
temperature mobility as high as $4.8\times 10^6$cm$^2$/Vs) we have measured
more than 20 ballistic conductance plateaus, with a high degree of flatness
that reflects the lack of potential fluctuations in the constriction. 
The samples show well defined 1D characteristics with little inter-subband
scattering, even between the closely spaced ($0.5$~meV) higher subbands. 
With a magnetic field applied in the plane of the 2DEG, each doubly
degenerate 1D subband is split by a Zeeman energy. Spectroscopy of the 1D
subbands can be performed\cite{patel91b} using a dc source-drain voltage
$V_{sd}$ and we have used this to measure the $g$-factors of the 1D
subbands.\cite{patel91a}

\section{Samples and Experimental Details}
\label{s:exp}

\subsection{Device Fabrication}
\label{s:MBE}
 
Measurements are presented here for six different samples, fabricated from
2DEGs formed in modulation-doped GaAs/Al$_{0.33}$Ga$_{0.67}$As
hetero\-structures, grown by molecular beam epitaxy (MBE) on a (100)
semi-insulating GaAs substrate. The sample properties are listed in Table~1.

For the single heterojunction samples (A-E), the 2DEG is formed at the
interface between a thick (1-2~$\mu$m) undoped GaAs buffer layer and a
600-1000~\AA\ undoped AlGaAs spacer layer. Doping is provided by 2000~\AA\ of
Si-doped AlGaAs ($1.2\times10^{17}$cm$^{-3}$), which is capped with 170~\AA\
of undoped GaAs. The use of lightly doped AlGaAs and a thick spacer layer
reduces the remote ionized impurity scattering and enhances the mobility. The
growth sequence for the quantum well sample (F) starts with a 100-period
25~\AA\ GaAs/AlGaAs superlattice buffer, which is used to trap surface
impurities from the substrate, and to progressively improve the interface
smoothness. This is followed by 1000~\AA\ of Al$_x$Ga$_{1-x}$As ramped from
$x=0$ to $x=0.33$, and a 0.45~$\mu$m buffer with $x=0.33$. Below the 200~\AA\
GaAs quantum well there is a 2000~\AA\ Si-doped Al$_{0.33}$Ga$_{0.67}$As
layer and an 800~\AA\ AlGaAs spacer, and above there is a 1000~\AA\ spacer
and a 400~\AA\ doped layer. The wafer is capped with 170~\AA\ of undoped
GaAs. 
On the back of all the wafers there is indium used to mount the
samples during MBE growth; this diffuses approximately $150$~\AA\ into the
GaAs substrate, and forms a back gate $350~\mu$m below the 2DEG. When the
back gate voltage $V_{bg}$ is changed from $-100$~V to $+50$~V, there is a
30\% increase of the carrier density $n_{2D}$.

The samples were first patterned into Hall bars. Ohmic contacts were made by
thermal evaporation of Au/Ge/Ni alloys, which were annealed for 80 seconds at
430$^{\circ}$C in a N$_2$/H$_2$ atmosphere. Split-gates were then patterned by
electron-beam lithography followed by thermal evaporation of 15~nm NiCr and
35~nm Au. All the split-gates had a length $L=0.4$~$\mu$m, with widths $W$
given in Table~1.

\subsection{Experimental Details}
\label{s:meas}

The two-terminal differential conductance of the samples, $G=dI/dV$, was
measured at low temperatures ($0.05-4$~K) in a dilution refrigerator using a
constant excitation voltage of 10~$\mu$V at 85~Hz. Measurements were also
performed with a high in-plane magnetic field ($B_{\parallel}$) applied
parallel to the current through the 1D constriction. To check for an
out-of-plane field component due to misalignment, we monitored the Hall
voltage; from such measurements we were able to align samples to better
than $1^{\circ}$. 

We use a technique developed by Patel {\it et al.}\cite{patel91a} to
deduce the energy separation of 1D subbands from the effects of an applied dc
source-drain voltage $V_{sd}$. A peak occurs in the transconductance
$dG/dV_g$ (obtained by numerical differentiation of the conductance) at the
gate voltage where there is a step in $G(V_g)$.
There is a crossing of adjacent transconductance peaks   
when $eV_{sd} = \Delta E_{N,N+1}$, 
where $\Delta E_{N,N+1}$ is the energy separation between the 
$N$ and $N+1$ subbands.\cite{patel91a} 
A doubling of the transconductance peaks can also be brought about 
using a strong in-plane magnetic field
to lift the spin degeneracy of the 1D subbands. 
The $g$-factor can be determined by comparing        
the voltage $V_{sd}$ required to produce 
the same amount of splitting as the magnetic field, 
and comparing the two energy scales\cite{patel91b}
\begin{equation}
eV_{sd}=2g_{\parallel}\mu_B B_{\parallel}S. 
\label{energy} 
\end{equation} 
This technique is valid if the transconductance peak splittings
are linear in both $B_{\parallel}$ and $V_{sd}$.

All conductance characteristics have been corrected for a series resistance
($R_S$) that is typically less than 2~k$\Omega$; this includes contributions
from the 2DEG, the contact resistances between the Ohmic contacts and the
2DEG, and the wires down the probe. Series resistance corrections have also
been applied to the source-drain measurements.

\section{Results}
\label{s:overall} 

\subsection{Zero-Field Conductance Characteristics}
\label{s:quantiz} 

Figure~\ref{f:1} shows the gate characteristics 
$G(V_g)$ of sample C at 60~mK. 
As the gate voltage $V_g$ is made negative the 2DEG beneath the
split-gates is depleted at $V_g=-0.9$~V, giving a sharp drop in the
conductance shown in the overall characteristics in Fig.~\ref{f:1}(b). Once
the 1D channel is defined, further decreases of $V_g$ narrow the channel
and reduce the carrier density in the vicinity of the constriction; as
a result the 1D subbands are depopulated and 
the conductance decreases in steps of $2e^2/h$. 
The constriction pinches off at $V_g=-5.75$~V,
when all the 1D subbands are depopulated.  
Overall, there are 25 well resolved
conductance plateaus; the last 15 are shown in the main figure, after
correction for a series resistance of $R_{S} = 703~\Omega$. The plateaus are
quantized at $N(2e^2/h)$ to within 1\% accuracy.

In addition to the usual quantized conductance plateaus, there is a structure
at $0.7(2e^2/h)$, seen in all samples. This is shown in Fig.~\ref{f:2} for two
devices, one based on a quantum well (sample F), and the other on a standard
heterojunction (sample D) measured at $T=1.5$~K. 
The 0.7 structure is not as precisely quantized 
as the conductance plateau at $2e^2/h$,
but is observed in the range $0.65 - 0.75(2e^2/h)$.
 
The 0.7 structure has distinctive dependences on carrier density and
temperature. Figure~\ref{f:3}(a) shows the gate characteristics $G(V_g)$ of
sample E for different 2D carrier densities. As $n_{2D}$ is decreased from 1.4
to $1.1\times10^{11}$cm$^{-2}$ using the back gate, the pinch-off voltage
becomes more positive. At the highest density, shown in the left hand trace,
the 0.7 structure is visible only as a weak knee in the gate characteristics,
which develops into a stronger structure as $n_{2D}$ is reduced.
 Figure~\ref{f:3}(b) shows the conductance $G(V_g)$ at $n_{2D} =
1.3\times10^{11}$cm$^{-2}$ as the temperature is raised from 0.1~K to 1.2~K in
steps of 0.1~K. The pinch-off voltage remains independent of temperature. The
plateau at $2e^2/h$ becomes thermally smeared at the highest temperature,
whereas the 0.7 structure becomes stronger, in agreement with previous
measurements\cite{thomas96a} of sample B. Figure~\ref{f:3}(c) shows the
temperature dependence at $n_{2D} = 1.0\times10^{11}$cm$^{-2}$; at this lower
electron density the more prominent 0.7 structure is less sensitive to
temperature. 
At higher temperatures, $T \sim 10$~K, the 0.7 structure disappears as does any
other subband feature. From this behavior, we tentatively ascribe to the
structure a characteristic energy of order 1~meV.

By applying different voltages to the two arms of the split-gate the 1D
channel can be moved laterally,\cite{stroh89,glazman91} allowing the
electrostatic potential landscape between the split-gates to be scanned.
Figure~\ref{f:4} shows the conductance characteristics obtained when the two
arms of the split-gate are swept together, but maintaining a constant voltage
difference $\Delta V_g$ between them. A change of $\Delta V_g$ from 0 to
1.3~V moves the channel by 80~nm; the plateau at $2e^2/h$ is unaffected by
the shift (as are the higher index plateaus) showing that the constriction is
free of impurities. In this sample the 0.7 structure occurs at
$0.65(2e^2/h)$, and is also unchanged by the lateral shift of the channel.
 
\subsection{Magnetic Field Dependence} 
\label{s:mag}

A strong in-plane magnetic field $B_{\parallel}$ 
lifts the spin degeneracy of the 1D subbands
giving conductance plateaus quantized in units of $e^2/h$.
Figure~\ref{f:5} shows the transconductance peaks in sample D  
split as $B_{\parallel}$ is increased in steps of 1~T.
As previously observed\cite{thomas96a} in sample A, there is an overall
parabolic shift of the gate characteristics with $B_{\parallel}$ that can be
attributed to a diamagnetic shift of both the 1D and 2D subband edges.\cite{smith88}
Satellite peaks, marked with an asterix 
($\ast$) and a solid bullet ($\bullet$),
corresponding to the conductance structures 
at $0.7(2e^2/h)$ and $1.7(2e^2/h)$, 
grow out of the right hand shoulders 
of the zero-field transconductance peaks. 
At the highest magnetic field, $B_{\parallel}=16$~T,
the transconductance peaks have roughly equal integrated areas, 
with the zeros between them corresponding to the
conductance plateaus quantized in units of $e^2/h$.
The Fig.~\ref{f:5} inset shows the voltage splitting 
$\delta V_g(B_{\parallel})$ for the first three subbands.
The Zeeman splittings are linear in $B_{\parallel}$, 
and at zero field the peak separation $\delta V_g(0)$ 
is finite for both $N=1$ and 2;
this demonstrates that the zero-field 0.7 structures evolve 
continuously into spin-split half-plateaus as the magnetic field is increased. 
By comparing $\delta V_g(0)$ to a 
$V_{sd}$-induced splitting, 
we estimate the zero-field energy gap as $\Delta_1 = 1.1$~meV 
for the lowest subband, and $\Delta_2 = 0.43$~meV for $N=2$. 
In our previous measurements\cite{thomas96a}
of sample~A we measured a zero-field gap $\Delta_1 = 1$~meV. 
In samples A and D the energy $\Delta_1$ is comparable 
to the temperature at which the 0.7 structure smears out. 

From the splitting of the transconductance peaks in $B_{\parallel}$
and $V_{sd}$, Eq.~\ref{energy} is used to determine the $g$-factors for all
1D subbands.\cite{note}
Figure~\ref{f:6} shows $g_{\parallel}$ measured as a function of
subband index $N$ for three different samples, as well as showing results for
sample~A at two different magnetic fields. 
When the channel is wide 
and there are many 1D subbands, 
the measured $g_{\parallel}$ is close to the bulk GaAs
value,\cite{white72} $\mid g \mid \approx 0.4$. 
As the number of occupied 1D
subbands decreases there is an enhancement of $g_{\parallel}$.

\subsection{The Effect of a Source-Drain Voltage $V_{sd}$} 
\label{s:sd}

The effect of a source-drain voltage $V_{sd}$ on the conductance 
characteristics $G(V_g)$ has been studied in detail in Ref.~\onlinecite{patel91b}.
As $V_{sd}$ is increased, half-plateaus appear at 
$(N + \frac{1}{2}) 2e^2/h$ for $G > 2e^2/h$, 
whereas $V_{sd}$-induced structures appear 
at $0.85(2e^2/h)$ and $0.3(2e^2/h)$ for $G < 2e^2/h$. 
% We focus here on the structure at $0.85(2e^2/h)$, 
% investigating it at two different temperatures. 
The gate voltage scale is a smooth measure of the 1D confinement energy, so a 
greyscale plot of the transconductance 
(similar to those presented in Ref.~\onlinecite{thomas95}) 
allows us to follow the energy shifts of subband features. 
Figure~\ref{f:7}(a) shows how the gate voltage positions of
transconductance features for the lowest three subbands move
as a function of $V_{sd}$ at $T=1.4$~K. 
The dark lines show transitions between plateaus and the 
white regions are the conductance plateaus (where the numbers 
denote the conductance in units of $2e^2/h$). 
Features moving to the right (left) with increasing $V_{sd}$
do so as the electrochemical potential of the 
source (drain) crosses a subband edge, 
and if the subband energies were independent of their occupation we
would expect a linear evolution of the transconductance 
structures with $V_{sd}$.
This is clearly not the case for the features associated 
with the 0.7 structure in the lowest subband,
suggesting that the subband configuration is occupation-dependent,
for which an interaction effect could be responsible. 
In Fig.~\ref{f:7}(b), we present data taken at $T=50$~mK, 
when the 0.7 structure is no longer visible at $V_{sd}=0$. 
As the electrochemical potential in the drain is
lowered below that of the source, 
a feature separates from the $N=1$ subband edge, 
giving rise to a white region that corresponds to a 
plateau at $0.85(2e^2/h)$.\cite{patel91b} 
Similar structures with smaller separations can be seen 
in Fig.~\ref{f:7}(b) for $N=2$ and $N=3$. 

\section{Discussion} 
\label{s:disc}        

\subsection{Evidence for a Spin Mechanism} 

In all the 18 samples from 7 different wafers that we have studied, the 0.7
structure is observed on all cool-downs. 
The structure has been measured in
both pointed\cite{VanWees91} and rectangular split-gates, 
in single heterojunction samples and in quantum wells, 
and is independent of the
distance of the 1D electron gas from the confining gate.\cite{thomas96a,patel91b} 
Recently, Kristensen {\it et
al.}\cite{krist97a} have observed a clear 0.7 structure in wires fabricated
by shallow etching, which provide stronger electrostatic confinement than
conventional split-gate structures. 
Some evidence of additional structure has
also been reported for GaAs wires\cite{tsch96} patterned by focussed ion beam
and InP based quantum wires,\cite{ramvall97} though in both cases the samples
are not of high mobility. 
We believe that the 0.7 structure is an intrinsic
property of clean 1D ballistic constrictions at low electron densities.

Additional structures in the gate characteristics of a ballistic 1D
constriction could be caused by impurity effects.\cite{McEuen90} 
Close to pinch-off the carrier density around the
constriction may become inhomogeneous,\cite{davies89} 
and the charging characteristics of small puddles of
electrons can give rise to Coulomb blockade peaks in the $G(V_g)$
characteristics.\cite{jtn93} 
Transmission resonances due to the multiple
reflection of electrons can also introduce conductance features below $2e^2/h$.
Our measurements cannot be explained by either of these mechanisms,
as both Coulomb blockade peaks and resonance phenomena undergo an
energy averaging at finite temperature which smears out their structure,
whereas Fig.~\ref{f:3} shows the 0.7 structure becoming stronger 
when the temperature is initially raised. 
It is also common that impurity effects differ
between sample cool-downs, which we do not observe. 
The clean quantized conductance plateaus
(see Fig.~\ref{f:1}) and the absence of additional structures when the channel is
moved from side to side (Fig.~\ref{f:4}) demonstrate the lack of potential
fluctuations in and around the 1D constriction in our samples. 

If the 0.7 structure were a transmission effect, unrelated to spin, it would be
replicated at $0.35(e^2/h)$ when the spin degeneracy was 
lifted by an applied magnetic field. 
This is not observed in the high field measurements shown in Fig.~\ref{f:5},
suggesting that the zero field intercepts of the spin splitting are related to
a spontaneous lifting of the spin degeneracy in the lowest subbands. 
A spontaneous spin polarization driven by an exchange interaction 
is predicted in a dilute 1D electron gas for both hard wall
cylindrical\cite{gold90} and parabolic\cite{gold96,wang96} confinements, and
the enhancement of the in-plane $g$-factor shown in Fig.~\ref{f:6} underlines
the importance of exchange effects as the 1D subbands are
depopulated.\cite{thomas96a,patel91a,daneshvar97} 
Figure~\ref{f:3}(a) shows that
the 0.7 structure strengthens as $n_{2D}$ is lowered, 
behavior that is consistent with an 
exchange interaction mechanism. 
Further evidence that an exchange mechanism may be responsible
for the 0.7 structure is provided by the source-drain measurements in Fig.~\ref{f:7}, 
where the features in the lowest subband are sensitive to the 
occupation statistics in the channel. 

Zero-field spin splitting could also arise from the spin-orbit
interaction, either from the inversion asymmetry of the conduction band of
GaAs, or internal electric fields due to the asymmetry of the confinement in
the growth direction. However, the energy of the spin-orbit term due to the
inversion asymmetry is calculated\cite{glazman89} to be only
$\sim 10^{-2}$~K, and measurements of the quantum well sample, see
Fig.~\ref{f:2}, show that the 0.7 structure is not weakened when the
confinement is less asymmetric.
%  Further, the strongly non-linear behavior in
% the greyscale plots of Fig.~\ref{f:7} show that the additional structures in
% the lowest subbands have energies which depend on the subband occupation, and
% are therefore associated with an electron-electron interaction.

Another mechanism for a spin polarization is based\cite{fasol94} on the
assumption that electron-electron scattering rates for hot electrons in a 1D
channel will be different for spin-up and spin-down electrons. However, the
0.7 structure is observed in equilibrium measurements ($V_{sd} = 0$) 
when there are no hot electrons. 
 
In summary, the 0.7 structure appears to be linked to spontaneous lifting of
spin degeneracy in the 1D constriction, driven by an electron-electron
interaction effect, and the evidence is initially consistent with this being the
exchange interaction. A spontaneous spin polarization of the electron gas,
however is expected to give rise to a conductance plateau at $0.5(2e^2/h)$,
rather than a structure at $0.7(2e^2/h)$. To address this point Wang and
Berggren\cite{wang98} propose that if the height of the saddle-point potential
is different for the two different spin orientations, then propagation of one
spin-split subband with some tunneling transmission probability for the other
spin may give a conductance above $0.5(2e^2/h)$. 
In an alternative theory, 
Schmeltzer {\it et al.}\cite{schmeltzer97} propose that within TL theory there
is a hybridization of the up and down spins in the last subband.

The temperature dependence of the 0.7 structure, where initially the feature
becomes stronger with increasing temperature, is also surprising;
a straightforward spin polarization is expected to weaken with increasing temperature. 
There is instead a characteristic temperature (1.5~K)
at which the 0.7 structure is most prominent,
and measurements\cite{krist97a} of the 
activated behavior of the 0.7 structure support this view.

\subsection{Relevance to the TL Model} 
\label{s:lutt}

As a consequence of electron-electron interactions, a 1D electron gas is
expected to exhibit Tomonaga-Luttinger\cite{tomanaga50,luttinger63} (TL)
liquid behavior. It is predicted\cite{kane92c} that the conductance of a
clean one-dimensional wire with a single conducting mode may be renormalized
to a value $K(2e^2/h)$, where $K>1$ for attractive interactions, and $K<1$
for repulsive interactions. It was later
argued\cite{maslov95a,safi95,pomarenko95} that such a conductance
renormalization may not occur, because the measured contact resistance is
determined by non-interacting electrons that are injected from the contacts
into the 1D wire. Impurity scattering, however, may give rise to corrections
to the low temperature dc conductance due to temperature and the finite
length of the system.\cite{ogata94,maslov95b} 

TL liquid behavior has been investigated in quantum wires fabricated by
two different techniques. Tarucha {\it et al.}\cite{tarucha95} fabricated
2-10~$\mu$m long 1D wires using wet etching and gating, and although no
renormalization of the conductance quantization was observed, the temperature
dependence of the last plateau is consistent with an interaction parameter
$K\approx0.7$ when fitted to a modified TL theory.\cite{ogata94} Using
cleaved edge overgrowth, Yacoby {\it et al.}\cite{yacoby96a} have fabricated
wires of length 1-20~$\mu$m that are strongly confined in both directions 
perpendicular to the wire axis. 
The wires have extremely high $L/W$ ratios, 
and clean conductance plateaus were observed, but quantized in units of
$\alpha(2e^2/h)$, where $0.75 < \alpha \leq 1$ 
and is both sample and temperature dependent.   
Recent theoretical work\cite{alekseev98} shows that 
these experimental results may be a consequence 
of enhanced backscattering at 
the interface between the 1D wire and the connecting 2DEG regions, 
and that the  nonuniversal quantization is not an intrinsic 
property of a one-dimensional electron gas. 
There is stronger evidence for TL behavior 
in the fractional quantum Hall regime.\cite{mill95,chang96}

We emphasize that our results are different from Yacoby {\it et al.}, in that we
observe a plateau at $2e^2/h$ {\em and} a structure at $0.7(2e^2/h)$, whereas
they observe non-universal quantization of the integer plateaus. Though our
results are not inconsistent with the TL as opposed to the Fermi liquid
description of the system, the effects which we present here are of a
different type, and relate to interactions between the two one-dimensional
liquids in opposite spin states. A model which includes spin-spin interactions
will therefore be necessary to adequately model our results.

\section{Conclusions} 
\label{s:conc}

In all the samples investigated, we observe clean quantized conductance
plateaus as well as the structure at $0.7(2e^2/h)$. We have shown that the
0.7 structure is not due to transmission or resonance effects, nor does it
arise from the asymmetry of the heterojunction in the growth direction. The
structure is not precisely quantized at $0.7(2e^2/h)$, and in a strong
in-plane magnetic field it moves to $0.5(2e^2/h)$. The origin of this 0.7
structure cannot be described by either Tomonaga-Luttinger theory or a simple
spin polarization of the electron gas, but we believe the exchange-enhanced
$g$-factor and the non-linear behavior of subband features with an
applied voltage to provide strong evidence that interaction effects are becoming
increasingly important as the 1D channel depopulates, and that the origin of the
0.7 structure is related to spin. 
 
\subsection*{Acknowledgements}

We thank the Engineering and Physical Sciences Research Council (UK) for
supporting this work. KJT acknowledges support from the Association of
Commonwealth Universities, JTN acknowledges an Advanced EPSRC Fellowship, and
DAR acknowledges support from Toshiba Cambridge Research Centre. We thank Drs
A.V. Khaetskii and C.H.W. Barnes for useful discussions. 

\end{multicols}

\begin{table}         
\caption{Sample properties} 
\begin{tabular}{|c|c|c|c|c|c|}

Sample\tablenote{Samples A and B were used
in Ref.~\onlinecite{thomas96a}.} & Structure\tablenote{SH=Single Heterojunction, 
QW=Quantum Well of width 200~\AA.} & 2DEG depth ($\mu$m) & 
Mobility\tablenote{The low temperature mobility $\mu$ and carrier density
$n_{2D}$ were measured at zero back gate voltage after illumination with a
red light-emitting diode.} $\mu$ ($10^6$cm$^2$/Vs)  & Carrier density
$n_{2D}$ ($10^{11}$cm$^{-2})$ & Split-gate\tablenote{All split-gates have a
length $L=0.4$~$\mu$m.} width $W$ ($\mu$m) \\ 
 \hline  
 A &  SH   & 0.28  & 4.5 & 1.8 & 0.75 \\  %T214A, PRL sample A  
 B &  SH   & 0.31  & 3.5  & 1.4 & 0.95 \\  %T207B, PRL sample B 
 C &  SH   & 0.28  & 4.5  & 1.8 & 0.95 \\  %T214B 
 D &  SH   & 0.31  & 3.5  & 1.4 & 0.75 \\  %T207A 
 E &  SH   & 0.29  & 3.5  & 1.3 & 0.75 \\  %T256 
 F &  QW   & 0.17  & 4.8  & 2.4 & 0.75 \\  %T258  
 \hline  
 \end{tabular} \end{table}

\begin{figure} 
\caption{The differential conductance $G(V_g)$ of sample C at $T=60$~mK,
after correction for a series resistance of $R_S = 703~\Omega$. Insets: (a)
Schematic of a split-gate device, where S and D represent the source and drain
contacts. (b) Raw data showing the definition and pinch-off
characteristics.} 
\label{f:1} 
\end{figure}

\begin{figure} 
\caption{Conductance characteristics of 1D constrictions defined in (a) a
quantum well, and (b) a conventional heterostructure.} 
\label{f:2} 
\end{figure}

\begin{figure}
\caption{(a) The 0.7 structure in sample E at 60~mK, as $n_{2D}$ is reduced
from $1.4\times10^{11}$cm$^{-2}$ ($V_{bg}=60$~V) to
$1.1\times10^{11}$cm$^{-2}$ ($V_{bg}=-110$~V) in steps of
$1.8\times10^{9}$cm$^{-2}$. The temperature dependence of the 0.7 structure,
in steps of 0.1~K, at (b) $n_{2D} = 1.3\times10^{11}$cm$^{-2}$ and (c)
$1.0\times10^{11}$cm$^{-2}$.} 
\label{f:3} 
\end{figure}

\begin{figure}
\caption{Lateral shifting of the channel in sample B at $T=60$~mK, using an
offset voltage $\Delta V_g$ between the two arms of the split-gate.
Each time $\Delta V_g$ is incremented by 0.1~V, 
the center of the 1D channel is shifted by 6.2~nm.} 
\label{f:4} 
\end{figure}

\begin{figure}
\caption{The transconductance $dG/dV_g$ of the first three subbands of
sample D as $B_{\parallel}$ is incremented in steps of 1~T. The right hand
shoulders, indicated with an asterix ($\ast$) and a solid bullet ($\bullet$),
show the features measured in the conductance at $0.7(2e^2/h)$ and
$1.7(2e^2/h)$ at $B_{\parallel}=0$~T. The inset shows the magnetic field
induced gate voltage splittings, $\delta V_g(B_{\parallel})$, for 
subband indices $N=1, 2$, and 3. 
The solid lines are least-squares linear fits to the data.} 
\label{f:5} 
\end{figure}

\begin{figure} 
\caption{The in-plane $g$-factor
$g_{\parallel}$ as a function of subband index $N$. The dashed line
at $\mid g \mid = 0.44$ indicates the $g$-factor for bulk GaAs.} 
\label{f:6} 
\end{figure}

\begin{figure} 
\caption{Greyscale plots of the zero field transconductance of sample F as a function of
$V_{sd}$ at (a) $T=1.2$~K, and (b) $T=50$~mK. 
The numbers indicate the plateau conductances in units of $2e^2/h$, and the 0.7
structure is the bright region at $V_{sd}=0$ between $G=0$  and $G=2e^2/h$ 
in the higher temperature data.
The usual linear splitting does not occur for features associated with the 0.7
structure, indicating that the energies of these features are
sensitive to the occupation of the subband. Note that similar features are
seen for $N=2$ and $N=3$.
The data in (a) and (b) were measured a week apart, 
over which time there was a slight change in the gate voltage 
characteristics of the device.} 
\label{f:7}  
\end{figure}

% \bibliography{/mnt2/JTN/tex/bib/Main}

\begin{thebibliography}{10}
\bibitem{note}
Where there is a zero-field splitting $\Delta_i$, we use the equation
  $eV_{sd}=2g_{\parallel}\mu_B B_{\parallel}S + \Delta_i$.
\end{thebibliography}
% \bibliographystyle{/mnt2/JTN/tex/BST/prstyjtn}

\end{document}